\documentclass[a4paper]{article}
\usepackage{graphicx}
\begin{document}

\begin{center}
{\bf {\Large Study of the neutron skin thickness 
of ${}^{208}$Pb in mean field models}}
\end{center}

\vspace{5mm}

\centerline{\rm X. Roca-Maza$^{1,2}$, M. Centelles$^1$, 
X. Vi\~nas$^1$ and M. Warda$^{1,3}$}

\vspace{2mm}

\begin{itemize}
\item[] $^1$Departament d'Estructura i Constituents de la Mat\`eria 
and Institut de Ci\`encies del Cosmos, Facultat de F\'{\i}sica, 
Universitat de Barcelona, Diagonal {\sl 647}, {\sl 08028} Barcelona, 
Spain.
\item[] $^2$INFN, sezione di Milano, via Celoria 16, I-{\sl 20133} 
Milano, Italy.
\item[] $^3$Katedra Fizyki Teoretycznej, Uniwersytet Marii 
Curie--Sk\l odowskiej, ul. Radziszewskiego 10, 20-031 Lublin, Poland.
\end{itemize}

\vspace{2mm}

\begin{abstract}
We study whether the neutron skin thickness $\Delta r_{np}$ of
$^{208}$Pb originates from the bulk or from the surface of the neutron
and proton density distributions in mean field models. We find that
the size of the bulk contribution to $\Delta r_{np}$ of $^{208}$Pb
strongly depends on the slope of the nuclear symmetry energy, while
the surface contribution does not. We note that most mean field models
predict a neutron density for $^{208}$Pb between the halo and skin
type limits. We investigate the dependence of parity-violating
electron scattering at the kinematics of the PREX experiment on the
shape of the nucleon densities predicted by the mean field models for
$^{208}$Pb. We find an approximate formula for the parity-violating
asymmetry in terms of the central radius and the surface
diffuseness of the nucleon densities of $^{208}$Pb in these models.
\end{abstract}

\section{Introduction}
\label{introduction}
The neutron skin thickness of nuclei is defined as the difference 
between the root mean square radius of neutrons and protons, 
\begin{equation}
\Delta r_{np}= \langle r^2 \rangle_n^{1/2} - \langle r^2 \rangle_p^{1/2}\ .
\label{rnp}
\end{equation}
The extraction of neutron radii and neutron skins from the experiment 
is in general dependent on the shape of the neutron distribution used 
in the analysis \cite{trz01,don09,fried09,cen10}. The data typically do 
not indicate unambiguosly if the difference between the peripheral neutron 
and proton densities is caused by an extended bulk radius of the neutron 
density, by a modification of the width of the surface, or by some 
combination of both effects. In particular, the neutron skin $\Delta r_{np}$ 
of $^{208}$Pb is nowadays attracting significant interest in both experiment 
and theory since it has a close relation with the density dependence of the 
nuclear symmetry energy and with the equation of state of neutron-rich matter 
\cite{bro00}, which have a large impact in diverse problems of nuclear physics 
and astrophysics where neutron-rich matter is involved \cite{bro00,cen09,cen09b}.

\section{Method}
\label{method}
The analysis of bulk and surface contributions to the neutron skin thickness 
of a nucleus requires proper definitions of these quantities based on nuclear 
density distributions \cite{warda10}. Using the standard definitions of the 
equivalent sharp radius $R$ and surface width parameter $b$ of a nucleon
density profile, we have shown in Refs.\ \cite{warda10,cen10} that one can 
write
$\Delta r_{np} = \Delta r_{np}^{\rm bulk} + \Delta r_{np}^{\rm surf}$,
where
\begin{equation}         
\label{rb}
\Delta r_{np}^{\rm bulk} \equiv
 \sqrt{\frac{3}{5}}\left(R_n-R_p\right)\nonumber
\end{equation}
and
\begin{equation}
\label{rs}
\Delta r_{np}^{\rm surf} \equiv
\sqrt{\frac{3}{5}} \, \frac{5}{2}
   \Big(\frac{b_n^2}{R_n}-\frac{b_p^2}{R_p}\Big) . \nonumber
\end{equation}
We recall that $R_q$ stands for the radius of a uniform sharp distribution 
whose density equals the bulk value of the actual density and that has the 
same number of particles. Therefore, one has a natural splitting of 
$\Delta r_{np}$ in terms of a bulk part (\ref{rb}) independent of surface 
properties, and a part (\ref{rs}) of surface origin \cite{warda10,cen10}.

\section{Results}
\label{results}

\subsection{Model predictions for bulk and surface contributions}
Nonrelativistic (NRMF) and relativistic (RMF) mean field models often differ 
in their predictions for properties of asymmetric nuclear matter. A common
example is the value predicted for the density slope $L$ of the nuclear 
symmetry energy at saturation density, which may show large discrepancies in
the MF interactions. The $L$ parameter is defined as 
\begin{equation}
L = \left. 3\rho_0\frac{\partial c_{\rm sym}(\rho)}{\partial \rho}
\right|_{\rho_0} \nonumber
\label{Lsym}
\end{equation}
where $c_{\rm sym}(\rho)$ is the symmetry energy and $\rho_0$ the nuclear 
saturation density. 
\begin{figure}
\begin{center}
\includegraphics[width=0.6\linewidth,angle=0,clip=true]{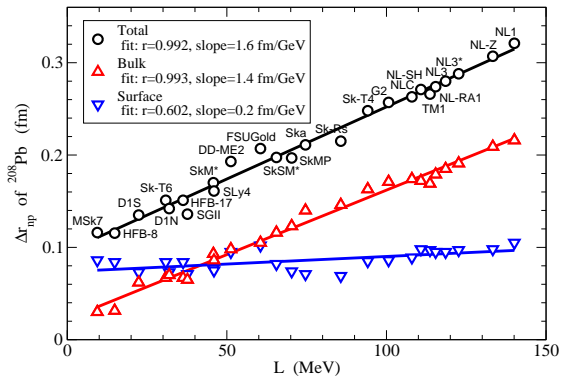}
\end{center}
\caption{\label{fig2} Linear correlation of $\Delta r_{np}$ of $^{208}$Pb with 
the density slope of the nuclear symmetry energy $L$. The dependence on $L$ 
of the bulk and surface contributions defined in Eqs.\ (\ref{rb}) and 
(\ref{rs}) is also displayed.}
\end{figure}

In Figure \ref{fig2} we show the linear correlation of $\Delta r_{np}$ of $^{208}$Pb 
with the parameter $L$ and demonstrate that it mainly arises from the bulk part
of $\Delta r_{np}$ within a large and representative set of mean field models of very
different nature. Relativistic (RMF) models: G2, NLC, NL-SH, TM1, NL-RA1, NL3, NL3*, 
NL-Z, NL1, DD-ME2 and FSUGold; and nonrelativistic (NRMF) models: HFB-8, MSk7, D1S, 
SGII, D1N, Sk-T6, HFB-17, SLy4, SkM*, SkSM*, SkMP, Ska, Sk-Rs and Sk-T4. The original 
references to the different interactions can be found in \cite{xu09,HFB-17} 
for the Skyrme models, \cite{cha08} for the Gogny models, and 
\cite{patra02b,sulaksono07,FSUG,DDME2,NL3S} and Ref.\ [19] in \cite{patra02b} 
for the RMF models. From this figure, one sees that whereas the bulk contribution 
to the neutron skin thickness of ${}^{208}$Pb changes largely among the different 
mean-field interactions, the surface contribution remains confined to a narrow band 
of values. The shape of the neutron density is more uncertain than the proton density 
in $^{208}$Pb, and even if the neutron rms radius is determined ({\it e.g.} in PREX 
\cite{prex2}), it can correspond to different shapes of the neutron density. As 
discussed in \cite{cen10}, from the study of the two-parameter Fermi functions fitted
to the self-consistent densities of the MF models, we find that for ${}^{208}$Pb the
central radii $C_q$ vary within the windows $C_n \approx 6.7-6.85$ fm in NRMF and 
$6.8-7$ fm in RMF, $C_p \approx 6.65-6.7$ fm in NRMF and $6.7-6.77$ fm in RMF, and that 
the diffuseness parameters $a_q$ vary within the windows $a_n \approx 0.53-0.55$ fm in 
NRMF and $0.55-0.59$ fm in RMF and $a_p \approx 0.43-0.47$ fm. From these results, 
we can conclude that most of the MF models predict a neutron distribution for $^{208}$Pb
between the halo-type limit (where $C_n-C_p=0$) and the skin-type limit (where $a_n-a_p=0$).
The halo-type is supported by models with a very soft symmetry energy. Models with a 
stiffer symmetry energy (larger $L$ values) have $C_n-C_p$ more and more different from 
zero. However, a pure skin-type is not found in any mean-field model as $a_n-a_p$ is 
always non-vanishing.

\subsection{Parity violating electron scattering at the kinematics of
the PREX experiment}

\begin{figure}
\begin{center}
\includegraphics[width=0.6\linewidth,angle=0,clip=true]{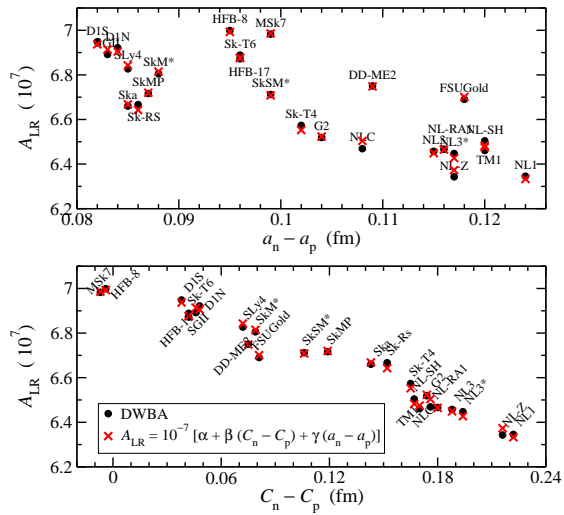}
\end{center}
\caption{\label{fig4} Parity violating asymmetry (DWBA) in $^{208}$Pb  
for 1~{\rm GeV} electrons at $5^\circ$ scattering angle. MF results and 
those obtained with the parametrized formula given in the text.}
\end{figure}

Parity-violating electron scattering (PVES) probes the neutron density in 
a nucleus since the $Z^0$ boson couples mainly to neutrons \cite{prex2}.
The PREX experiment at JLab \cite{prex1} aims to determine the neutron
rms radius of $^{208}$Pb by PVES. We have investigated the dependence of 
PVES on the shape of the neutron and proton densities of $^{208}$Pb within 
the MF models. To this end, we compute the parity-violating asymmetry
\begin{equation}
A_{LR}\equiv
\Big(\frac{d\sigma_+}{d\Omega}-\frac{d\sigma_-}{d\Omega}\Big) \Big/
\Big(\frac{d\sigma_+}{d\Omega}+\frac{d\sigma_-}{d\Omega}\Big)
\nonumber
\label{alr}
\end{equation}
at the PREX kinematics \cite{prex1}. We obtain the differentical cross
sections  $d\sigma_\pm/d\Omega$ for right and left-handed electrons by 
performing the exact phase shift analysis (DBWA) of the Dirac equation 
\cite{roca08} for incident electrons moving in the potentials 
$V_\pm(r)=V_{\rm Coulomb}(r) \pm V_{\rm weak}(r)$ \cite{cen10}. See 
Ref.~\cite{cen10} for the details of the calculations. We display in 
Fig.~\ref{fig4} the results for $A_{LR}$ as a function of $C_n-C_p$ and 
$a_n-a_p$ of the two parameter Fermi distributions fitted to the $^{208}$Pb 
MF densities. We have found that the results can be reasonably parametrized 
by the formula $10^7 A_{LR} \approx \alpha + \beta (C_n-C_p) + \gamma (a_n-a_p)$ 
with $\alpha=7.33$, $\beta=-2.45$ fm$^{-1}$ and $\gamma=-3.62$ fm$^{-1}$,
which is depicted by the crosses in the figure. Recently, in
\cite{arxiv11} we have analyzed systematically the correlations of
$A_{LR}$ with the neutron skin thickness of $^{208}$Pb and with the
slope of the symmetry energy in the nuclear MF models.

\section{Conclusions}
\label{conclusions}
We have found that the known linear correlation of $\Delta r_{np}$ of $^{208}$Pb 
with the density derivative of the nuclear symmetry energy arises from the bulk 
part of $\Delta r_{np}$. This implies that an experimental determination of $R_n$ 
of $^{208}$Pb could be as useful for constraining $L$ as a determination of 
$\langle r^2\rangle_n^{1/2}$. MF models can accomodate the halo-type distribution 
in $^{208}$Pb if the symmetry energy is very soft but do not support a purely 
skin-type distribution. We find a simple parametrization of $A_{LR}$ in terms of 
$C_n-C_p$ and $a_n-a_p$ that would provide a new correlation between the central
radius and the surface diffuseness of the distribution of neutrons in
$^{208}$Pb assuming the proton density known from experiment.

\section*{Acknowledgments}
Work partially supported by grants CSD2007-00042,
FIS2008-01661 and 2009SGR-1289 (Spain) and N202~231137 (Poland).

\end{document}